\documentclass[superscriptaddress,amsmath, nofootinbib]{revtex4}
\usepackage{amsfonts}
\usepackage{graphicx}
\usepackage[brazil]{babel}
\usepackage[latin1]{inputenc}
\usepackage{amsmath}
\usepackage{amssymb}
\begin{document}


\title{Euler-Poisson equations of a dancing spinning top, integrability and examples of analytical solutions.}

\author{Alexei A. Deriglazov }
\email{alexei.deriglazov@ufjf.br} \affiliation{Depto. de Matem\'atica, ICE, Universidade Federal de Juiz de Fora,
MG, Brazil} 

\author{}

\date{\today}

\begin{abstract}
Equations of a rotating body with one point constrained to move freely on a plane (dancing top) are deduced from the Lagrangian variational problem. They formally look like the Euler-Poisson equations of a heavy body with fixed point, immersed in a fictitious gravity field.  Using this analogy, we have found examples of analytical solutions for the case of a heavy symmetrical dancing top. They describe the motions with center of mass keeping its height fixed above the supporting plane. General solution to equations of a dancing top in terms of exponential of Hamiltonian field is given. 
An extra constraint, that take into account the reaction of supporting plane, leads to modification of the canonical Poisson structure and therefore the integrability according to Liouville is under the question.
\end{abstract}

\maketitle 



\section{Introduction.}
The dynamics of a rotating body can be formulated by choosing, as the only starting point, the variational problem for a system of particles subject to kinematic constraints. This approach allows one to describe a free rigid body as well as a body with one fixed point in a gravitational 
field \cite{AAD23,AAD23_9}. Here we show that this scheme equally works for a more complicated case of a body with one point constrained to move freely on a plane. It will be called the dancing top. 
In Sect. II we deduce equations of the dancing top and observe that they formally look like the Euler-Poisson equations of a heavy body with fixed point, immersed in a fictitious gravity field arising due to reaction of the supporting plane. However, the acceleration of the fictitious gravity turns out to be a complicated function of the rotational degrees of freedom. Using this observation, in Sect. III we present an example of analytic solution (see Eqs. (\ref{dt47}) and (\ref{dt51}) below), that describe 
the motion of a symmetric dancing top with center of mass keeping its height fixed above the supporting plane. This turn out to be  two-frequency motion with surprisingly reach relationship between the frequences, that, besides the inclination, depends also on the top's geometry. In Sect. IV we present the general solution to equations of a dancing top in terms of Hamiltonian vector field and then discuss the integrability of a symmetric dancing top according to Liouville. The Hamiltonian constraints, arising due to reaction of the supporting plane, modify the canonical Poisson brackets of the theory, and therefore the Liouville's integrability turn out to be under the question.

Being one of the classical problems in the theory of integrable systems and classical mechanics, these issues however could be of interest in the modern studies of various aspects related with construction and behavior of spinning particles and rotating bodies in external fields beyond the pole-dipole approximation \cite{Dey_2023,Fri_2023,Ram_2022,Zha_2022,Ion_2023,Tao_2023,Yas_2023,
Wan_2023,San_2021,Yun_2021,Yon_2022, Abd_2022,Dmi_2023,Dmi_2022,Kos_2023}.

\section{Heavy body with one point confined to move freely on a plane.}\label{FP}

Consider a rigid body (composed of $n$ points with the coordinates $y_N^i$, N=1, 2, \ldots , n, i=1, 2, 3),  one point of which is forced to move freely on a plane. To be specific, imagine a conical top whose point of base (fulcrum), say ${\bf y}_1(t)$, always moves on a horizontal plane, say $\pi$, without friction. Following the same procedure as in \cite{AAD23,AAD23_9}, our aim here will be to deduce equations of the top starting from a Lagrangian variational problem\footnote{We use the notation adopted in these works. In particular, 
the notation for the scalar product is:  $({\bf a}, {\bf b})=a_i b_i$. Notation for the vector product: $[{\bf a}, {\bf b}]_i=\epsilon_{ijk}a_j b_k$, 
where $\epsilon_{ijk}$ is Levi-Chivita symbol in three dimensions, with $\epsilon_{123}=+1$.}. We will show that equations of a dancing spinning top formally look like the Euler-Poisson equations of a heavy body with fixed point, immersed in a fictitious gravity field arising due to reaction of supporting 
plane $\pi$.  

Let ${\bf k}$ is unit normal vector to the plane $\pi$.  Assuming that the origin of Laboratory was chosen on this plane, the constraints of the problem are 
\begin{eqnarray}\label{dt1}
({\bf y}_N(t)-{\bf y}_K(t),  {\bf y}_P(t)-{\bf y}_M(t))=\mbox{const}, \qquad ({\bf k}, {\bf y}_1(t))=0.   
\end{eqnarray}
The orthogonal axes of the Laboratory are so far arbitrarily directed. Their direction will be fixed below so that the inertia tensor in equations of motion becomes a diagonal matrix (see the discussion below Eq. (\ref{dt22})).

We choose three points ${\bf y}_2(t), {\bf y}_3(t), {\bf y}_4(t)$ of the body so that ${\bf y}_1(t), {\bf y}_2(t), {\bf y}_3(t), {\bf y}_4(t)$ do not lie in the same plane. Then we can take $3(N-2)+1$ relations $({\bf y}_A(t)-{\bf y}_1(t),  {\bf y}_N(t)-{\bf y}_1(t))=\mbox{const}$,  $({\bf k}, {\bf y}_1(t))=0$ as an independent constraints among (\ref{dt1}). Here $A=2, 3, 4$. Then the Lagrangian which take into account these constraints is 
\begin{eqnarray}\label{dt2}
L_0=\frac12\sum_{N=1}^{n}m_N\dot{\bf y}_N^2+\frac12\sum_{A, B=2}^{4}\lambda_{AB}\left[({\bf y}_A-{\bf y}_1, {\bf y}_B-{\bf y}_1)-a_{AB}\right]+
\sum_{A=2}^{4}\sum_{\beta=5}^{n}\lambda_{A\beta}\left[({\bf y}_A-{\bf y}_1, {\bf y}_\beta-{\bf y}_1)-a_{A\beta}\right]+\sigma({\bf k}, {\bf y}_1), 
\end{eqnarray}
where the matrix $\lambda_{AB}$ was chosen to be symmetric, and $\sigma$ is the Lagrangian multiplier for the constraint $({\bf k}, {\bf y}_1)=0$. 
The force of gravity, with the acceleration of gravity equal to $a>0$, is taken into account adding the potential energy  
\begin{eqnarray}\label{dt3}
L=L_0-U, \qquad U=\sum_{N=1}^{n}am_N({\bf k}, {\bf y}_N). 
\end{eqnarray}
The variation of $L$ with respect 
to $\lambda_{AB}, \lambda_{A\beta}$ and $\sigma$ imply the constraints (\ref{dt1}), while the variation with respect to ${\bf y}_N(t)$ give the dynamical equations
\begin{eqnarray}\label{dt4}
m_1\ddot y^i_1=-\sum_{AB}\lambda_{AB}[y^i_B-y^i_1]-\frac12\sum_{A\alpha}\lambda_{A\alpha}[y^i_A+y^i_\alpha-2y^i_1]+\sigma k^i-am_1k^i, \cr
m_A\ddot y^i_A=\sum_{B}\lambda_{AB}[y^i_B-y^i_1]+\frac12\sum_{\alpha}\lambda_{A\alpha}[y^i_\alpha-y^i_1]-am_Ak^i, \cr
m_\alpha\ddot y^i_\alpha=\frac12\sum_{A}\lambda_{A\alpha}[y^i_A-y^i_1]-am_\alpha k^i. 
\end{eqnarray}

{\bf The center-of-mass variable.}   Taking the sum of equations (\ref{dt4}), we obtain $\sum_{N=1}^{n}m_N\ddot{\bf y}_N=-a(\sum_{N=1}^{n}m_N)k^i+\sigma k^i$. It is convenient to introduce the center of mass of the body as follows:
\begin{eqnarray}\label{dt5}
{\bf y}_c(t)=\frac1\mu\sum_{N=1}^{n}m_N{\bf y}_N(t), ~~\mbox{where} ~ \mu=\sum_{N=1}^{n}m_N, ~ \mbox{then} ~ 
\ddot{\bf y}_c=-(a-\frac{\sigma}{\mu}){\bf k}.
\end{eqnarray}
According to Eq. (\ref{dt5}), the part of force acting on the center of mass, arising due to freezing the coordinate ${\bf y}_1$ on the plane, looks like an extra gravity (when $\sigma<0$), with the acceleration equal to $-\sigma/\mu$. 
Using the projectors $N_{ij}({\bf k})\equiv \delta_{ij}-k_i k_j$, $P_{ij}({\bf k})\equiv k_i k_j$, we can decompose the vector of position ${\bf y}_c(t)$ on  orthogonal and longitudinal parts with respect to constant vector ${\bf k}$ 
\begin{eqnarray}\label{dt6}
{\bf y}_c(t)=N({\bf k}){\bf y}_c(t)+P({\bf k}){\bf y}_c(t)\equiv{\bf y}_{\bot c}(t)+{\bf y}_{\| c}(t).
\end{eqnarray}
The dynamics of these vectors is as follows:
\begin{eqnarray}\label{dt7}
\ddot{\bf y}_{\bot c}=0, \qquad \ddot{\bf y}_{\| c}=-(a-\frac{\sigma}{\mu}){\bf k}, 
\end{eqnarray}
that is the projection of center-of-mass onto the plane $\pi$ moves uniformly along a straight line, while the altitude of center-of-mass is subject to shifted  acceleration of gravity equal to $a-\frac{\sigma}{\mu}$. 

\begin{figure}[t] \centering
\includegraphics[width=09cm]{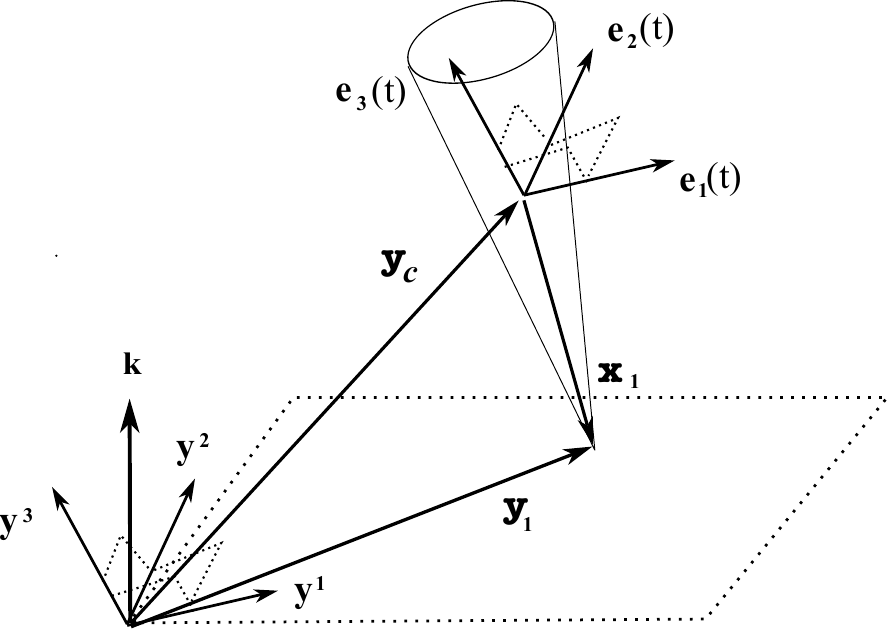}
\caption{Dancing spinning top with center of mass at the point ${\bf y}_c$. ${\bf x}_1$ is position vector of the fulcrum ${\bf y}_1$ with respect to the center of mass.}\label{DT_1}
\end{figure}

It is convenient to make a change of variables in the Lagrangian $L$, such that the position vector of the center-of-mass becomes one of the coordinates of the problem
\begin{eqnarray}\label{dt8}
({\bf y}_1, {\bf y}_2, \ldots , {\bf y}_n ) \rightarrow\left({\bf y}_c=\frac1\mu\sum_{N=1}^{n}m_N{\bf y}_N, ~ {\bf x}_P={\bf y}_P-{\bf y}_c\right), \quad P=1, 2, \ldots , n-1. 
\end{eqnarray}
The inverse change is
\begin{eqnarray}\label{dt9}
({\bf y}_c,  {\bf x}_1, \ldots , {\bf x}_{n-1}  )\rightarrow \left({\bf y}_P={\bf y}_c+{\bf x}_P,   ~ {\bf y}_n={\bf y}_c-\frac{1}{m_n}\sum_{1}^{n-1}m_N{\bf x}_N\right), \quad P=1, 2, \ldots , n-1. 
\end{eqnarray}
The variables ${\bf x}_P$ are the position vectors of $n-1$ points of the body with respect to the point of center of mass, see Figure \ref{DT_1}. 

An invertible change of variables can be performed in a Lagrangian, this is known to give an equivalent formulation of the initial problem \cite{Arn_1, Landau_8, deriglazov2010classical}. In terms of new variables (\ref{dt9}), our Lagrangian reads as follows:
\begin{eqnarray}\label{dt10}
L=\frac12 \mu\dot{\bf y}_c^2+\frac12\sum_{N=1}^{n}m_N\dot{\bf x}_N^2+\frac12\sum_{A=2}^{4}\sum_{N=2}^{n}\lambda_{AN}\left[({\bf x}_A-{\bf x}_1, {\bf x}_N-{\bf x}_1)-a_{AN}\right]+\sigma({\bf k}, {\bf y}_c+{\bf x}_1)-a\mu({\bf k}, {\bf y}_c),  
\end{eqnarray}
where it was denoted 
\begin{eqnarray}\label{dt11}
{\bf x}_n\equiv-\frac{1}{m_n}\sum_{1}^{n-1}m_N{\bf x}_N,  ~~\mbox{or} ~ \sum_1^n m_N{\bf x}_N=0. 
\end{eqnarray}
This prompts to introduce an independent auxiliary variable ${\bf x}_n$, and to take into account the equality (\ref{dt11}) as one more constraint of the problem, adding it to our action with the corresponding Lagrangian multiplier. According to the results known from classical mechanics, this gives an equivalent Lagrangian
\begin{eqnarray}\label{dt12}
L=\frac12 \mu\dot{\bf y}_c^2+\frac12\sum_{N=1}^{n}m_N\dot{\bf x}_N^2+\frac12\sum_{A, B=2}^{4}\lambda_{AB}\left[({\bf x}_A-{\bf x}_1, {\bf x}_B-{\bf x}_1)-a_{AB}\right]+\qquad \qquad \cr \frac12\sum_{A=2}^{4}\sum_{\alpha=5}^{n}\lambda_{A\alpha}\left[({\bf x}_A-{\bf x}_1, {\bf x}_\alpha-{\bf x}_1)-a_{A\alpha}\right] +\sigma({\bf k}, {\bf y}_c+{\bf x}_1)-a\mu({\bf k}, {\bf y}_c)+{\boldsymbol\lambda}[\sum_1^n m_N{\bf x}_N].  
\end{eqnarray}
It now consist of $3(n+1)$ independent dynamical variables ${\bf y}_0$ and ${\bf x}_N$, $N=1, 2, \ldots , n$, as well as $3n-2$ auxiliary variables $\lambda_{AN}$, $\boldsymbol{\lambda}$ and $\sigma$.  This action implies the following dynamical equations: 
\begin{eqnarray}\label{dt13}
m_1\ddot x^i_1=-\sum_{AB}\lambda_{AB}[x^i_B-x^i_1]-\frac12\sum_{A\alpha}\lambda_{A\alpha}[x^i_A+x^i_\alpha-2x^i_1]+m_1\lambda^i+\sigma k^i, \cr
m_A\ddot x^i_A=\sum_{B}\lambda_{AB}[x^i_B-x^i_1]+\frac12\sum_{\alpha}\lambda_{A\alpha}[x^i_\alpha-x^i_1]+m_A\lambda^i, \cr
m_\alpha\ddot x^i_\alpha=\frac12\sum_{A}\lambda_{A\alpha}[x^i_A-x^i_1]+m_\alpha\lambda^i, \cr
\mu\ddot y_c^i=\sigma k^i-a\mu k^i.  
\end{eqnarray}
They are accompanied by the constraints, arising from variation of the action with respect to Lagrangian multipliers 
\begin{eqnarray}\label{dt14}  
({\bf x}_A-{\bf x}_1, {\bf x}_N-{\bf x}_1)=a_{AN}, \qquad ({\bf k}, {\bf y}_c+{\bf x}_1)=0,  \qquad \sum_{1}^n m_N{\bf x}_N=0.  
\end{eqnarray}
Note that in equations (\ref{dt13}) gravity acts only on the center-of-mass point.

\noindent {\bf Action functional and second-order equations for the rotation matrix.} The new variables $x^i_N(t)$ have simple physical meaning. Consider (non inertial) coordinate system with origin at ${\bf y}_c(t)$ and coordinate axes in the direction of Laboratory axes for all $t$, see Figure \ref{DT_1}. Then $x^i_N(t)=y^i_N(t)-y^i_c(t)$ are coordinates of the body's point $N$ with respect to this center-of-mass system. Unit vectors along the axes of the system will be denoted by ${\bf e}_1, {\bf e}_2, {\bf e}_3$. Their components in the center-of-mass system are ${\bf e}_1=(1, 0, 0)^T$, ${\bf e}_2=(0, 1, 0, )^T$, ${\bf e}_3=(0, 0, 1)^T$. The lengths of ${\bf x}_N(t)$ and angles between them do not depend on time \cite{AAD23}. So their evolution is given by orthogonal time-dependent transformation 
\begin{eqnarray}\label{dt19}
{\bf x}_N(t)=R(t){\bf x}_N(0), \qquad \mbox{this implies} \qquad  R(0)={\bf 1},
\end{eqnarray}
or, equivalently, 
\begin{eqnarray}\label{dt19.1}
{\bf y}_N(t)={\bf y}_c(t)+R(t){\bf x}_N(0).  
\end{eqnarray}
The latter expression shows how the motion of the body's points can be restored knowing the evolution of center-of-mas ${\bf y}_c(t)$ and the rotation matrix $R(t)$. To obtain equations for $R_{ij}$, we substitute (\ref{dt19}) into (\ref{dt13}). Multipluing the resulting equation with number $N$ by $x_N^j(0)$  and taking their sum, we get 
\begin{eqnarray}\label{dt20}
\ddot R_{ik}g_{kj}=-R_{ik}\lambda_{kj}+\sigma k^i x_1^j(0), 
\end{eqnarray}
where it was denoted
\begin{eqnarray}\label{dt21}
g_{kj}\equiv\sum_{N=1}^{n} m_Nx_N^k(0) x_N^j(0),
\end{eqnarray}
and $\lambda_{jk}$ is a symmetric matrix composed of $\lambda_{AN}$ (we do not need its explicit form).  According to the physical meaning of $x^i_N(t)$ discussed above, $g_{kj}$ is the mass matrix calculated in the center-of-mass system at $t=0$.

The equations (\ref{dt20}) follow from their own variational problem
\begin{eqnarray}\label{dt22}
S=\int dt ~ ~ \frac12 g_{ij}\dot R_{ki}\dot R_{kj} -\frac12 \lambda_{ij}\left[R_{ki}R_{kj}-\delta_{ij}\right]+\frac12\mu\dot{\bf y}_c^2+\sigma({\bf k}, {\bf y}_c+R{\bf z})-a\mu({\bf k}, {\bf y}_c), \qquad \mbox{where} \quad {\bf z}\equiv{\bf x}_1(0).  
\end{eqnarray}
Besides, this implies the desired dynamical equation $\mu\ddot{\bf y}_c=-a\mu{\bf k}+\sigma{\bf k}$, as well as  the constraints $R^TR={\bf 1}$ and $({\bf k}, {\bf y}_c+R{\bf z})=0$  as the conditions of extremum of this variational problem. 

\begin{figure}[t] \centering
\includegraphics[width=16cm]{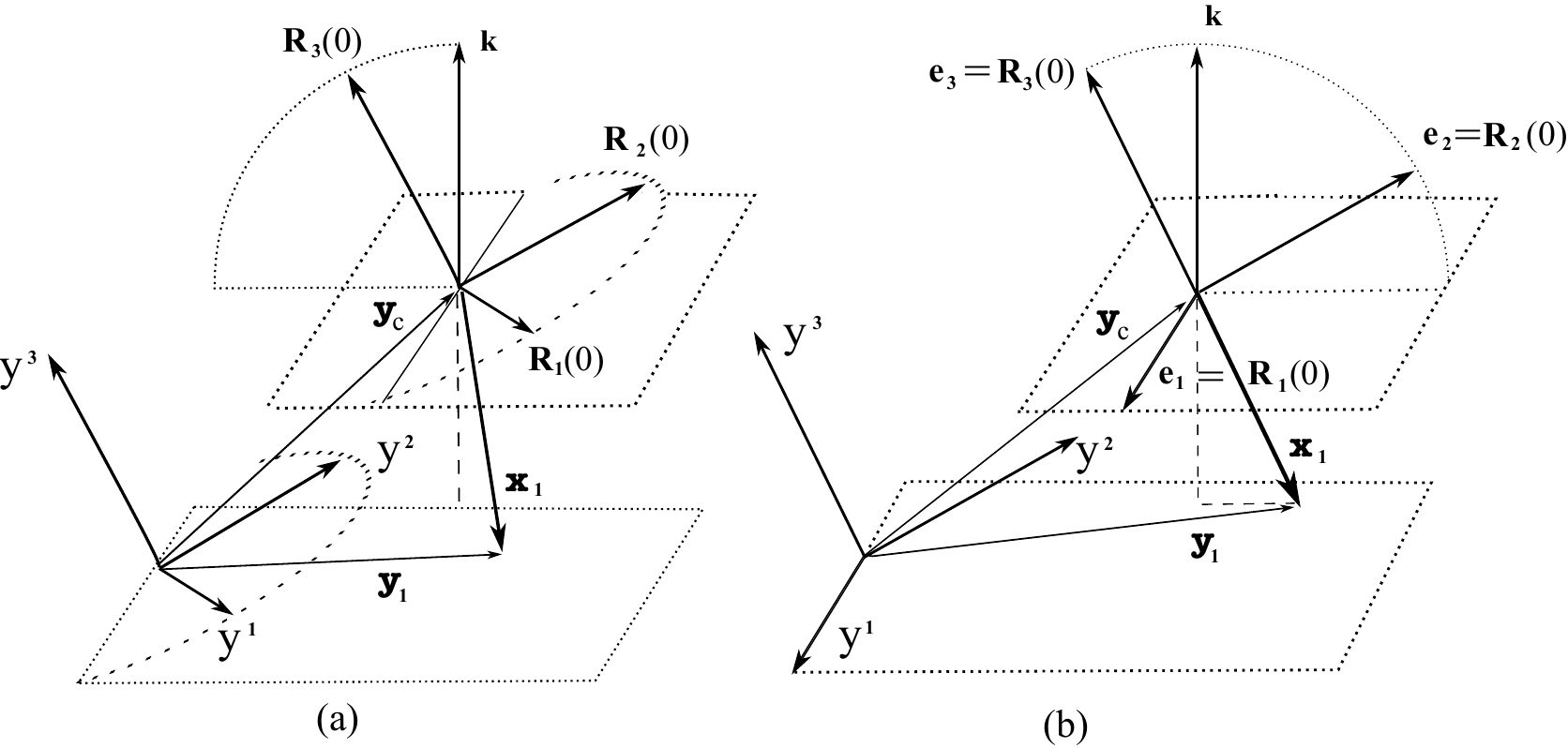}
\caption{(a) At $t=0$ the Laboratory axes $y_i$ should be chosen in the direction of principal axes of inertia ${\bf R}_i(0)$. (b) Symmetric 
top $I_1=I_2\ne I_3$ with a fulcrum on the third axis of inertia. Then without spoiling a diagonal form of the inertia tensor, the Laboratory system can be chosen such that the vectors ${\bf k}$, ${\bf R}_3(0)$ and ${\bf x}_1(0)$ lie on the plane of the sheet of paper. }\label{DT_2}
\end{figure}

Let ${\bf R}_i(t)$ be othonormal vectors with the origin at ${\bf y}_c(t)$, chosen in the directions of eigenvectors of $g_{ij}$ at $t=0$, and then rigidly connected with our body. Without loss of generality, we assume that at the instant $t=0$ the Laboratory axes $y^i$ were chosen in the directions of ${\bf R}_i(0)$, see Figure \ref{DT_2} (a) (in this figure, the vectors ${\bf k}$ and ${\bf R}_3(0)$ are chosen to lie in the plane of the sheet of paper). By construction, the axes $x^i$ of center-of-mass system also coincide with ${\bf R}_i(0)$. Then the mass matrix in the action  (\ref{dt22}) has the diagonal form: $g_{ij}=diagonal ~ (g_1, g_2, g_3)$. The corresponding inertia tensor $I_{ij}=\left[g_{kk}\delta_{ij}-g_{ij}\right]$ will be diagonal 
also\footnote{Failure to take this circumstance into account leads to a lot of confusion, see \cite{AAD23_3}.}: $I_{ij}=diagonal ~ (I_1, I_2, I_3)$. As the ends of the vectors ${\bf R}_i(t)$ represent the points rigidly connected to the body, say ${\bf x}_i(t)$, we can write according to (\ref{dt19}): $({\bf R}_i(t))_j=({\bf x}_i(t))_j=(R(t){\bf x}_i(0))_j=R_{jk}x_i^k(0)=R_{jk}\delta_i^k=R_{ji}$. That is ${\bf R}_i(t)$ coincide with columns of the rotation matrix, that was introduced above in an independent manner: $R(t)=({\bf R}_1(t), {\bf R}_2(t), {\bf R}_3(t))$. 
%
%


{\bf Exclusion of auxiliary variables and first-order equations for $R_{ij}$.} Assuming the diagonal form of $g_{ij}$, the equations (\ref{dt20}) read as follows (there is no summation over $i$ in Eq. (\ref{dt23}) ) 
\begin{eqnarray}\label{dt23}
g_i\ddot{\bf R}_i=-\sum_j\lambda_{ij}{\bf R}_j+\sigma{\bf k} z_i. 
\end{eqnarray}
To exclude the auxiliary variables $\lambda_{ij}$ from these equations, we use the othogonality constraint 
\begin{eqnarray}\label{dt24}
({\bf R}_i, {\bf R}_j)=\delta_{ij}.
\end{eqnarray}
Calculating its first and second derivatives we get its consequences 
\begin{eqnarray}\label{dt25}
(\dot{\bf R}_i, {\bf R}_j)+({\bf R}_i, \dot{\bf R}_j)=0, \qquad (\ddot{\bf R}_i, {\bf R}_j)+({\bf R}_i, \ddot{\bf R}_j)+2(\dot{\bf R}_i, \dot{\bf R}_j)=0. 
\end{eqnarray}
Using Eqs. (\ref{dt23}) for second derivatives in the last expression, the orthogonality constraint and the symmetry of $\lambda_{ij}$, we get after some calculations 
\begin{eqnarray}\label{dt26}
\lambda_{ij}=\frac{2g_i g_j}{g_i+g_j}(\dot{\bf R}_i, \dot{\bf R}_j)+\sigma\frac{g_i({\bf k}, {\bf R}_i)z_j+g_j({\bf k}, {\bf R}_j)z_i}{g_i+g_j}. 
\end{eqnarray}
Using these $\lambda_{ij}$  in (\ref{dt23}), we obtain equations that contain only one auxiliary variable $\sigma$. 
\begin{eqnarray}\label{dt27}
\ddot{\bf R}_i=-\sum_j\frac{2g_j}{g_i+g_j}(\dot{\bf R}_i, \dot{\bf R}_j){\bf R}_j-\sigma\frac{({\bf k}, {\bf R}_i)z_j-({\bf k}, {\bf R}_j)z_i}{g_i+g_j}{\bf R}_j. 
\end{eqnarray}
The last term contains an antisymmetric matrix, that can be replaced on an equivalent vector as follows:
\begin{eqnarray}\label{dt28}
\hat a_{ij}\equiv\frac{({\bf k}, {\bf R}_i)z_j-({\bf k}, {\bf R}_j)z_i}{g_i+g_j}=\epsilon_{ijk}a_k, \quad \mbox{where} \quad 
a_k=\frac12\epsilon_{kij}\hat a_{ij}=(I^{-1}[R^T{\bf k}, {\bf z}])_k, 
\end{eqnarray}
where $I_{ij}=diagonal (I_1, I_2, I_3)$ is the inertia tensor calculated in the center-of-mass system at $t=0$. 
Then Eq. (\ref{dt27}) acquires the form
\begin{eqnarray}\label{dt29}
\ddot{\bf R}_i=-\sum_j\frac{2g_j}{g_i+g_j}(\dot{\bf R}_i, \dot{\bf R}_j){\bf R}_j-\sigma\epsilon_{ijk}{\bf R}_ja_k. 
\end{eqnarray}
To exclude the auxiliary variable $\sigma$, we could use this equation in the expression $({\bf k}, \ddot{\bf y}_c+\ddot R{\bf z})=0$ to represent $\sigma$ through the physical variables $R_{ij}$ and $y^i_c$. However, it is more convenient first to rewrite equations (\ref{dt29}) in a first-order form. 


To this aim, consider the space of {\it mutually independent} dynamical variables $R_{ij}(t)$ and $\Omega_i(t)$, subject to the 
equations $({\bf R}_i, {\bf R}_j)=\delta_{ij}$, as well as to
\begin{eqnarray}
\ddot R_{aj}=-\sum_k\frac{2g_k}{g_j+g_k}(\dot{\bf R}_j, \dot{\bf R}_k)R_{ak}-\sigma\epsilon_{jnk}R_{an}a_k, \label{dt30} \\  
\Omega_k=-\frac12\epsilon_{kij}(R^T\dot R)_{ij}. \label{dt31}
\end{eqnarray}
That is $R_{aj}(t)$ satisfies the equations (\ref{dt29}), while $\Omega_k(t)$ accompanies the dynamics of $R_{aj}(t)$ according to (\ref{dt31}).  Evidently, this system is equivalent to (\ref{dt29}). Multiplying Eq. (\ref{dt30}) on the invertible matrix $R_{ai}$ we get
\begin{eqnarray}\label{dt32}
({\bf R}_i, \ddot{\bf R}_j)=-\frac{2g_i}{g_i+g_j}(\dot{\bf R}_i, \dot{\bf R}_j)+\sigma\epsilon_{ijk}a_k. 
\end{eqnarray}
Let us separate symmetric and antisymmetric parts of Eq. (\ref{dt32}) as follows:
\begin{eqnarray}
({\bf R}_i, \ddot{\bf R}_{j})+({\bf R}_j, \ddot{\bf R}_{i}) =-2(\dot{\bf R}_i, \dot{\bf R}_j), \label{dt33} \\ 
({\bf R}_i, \ddot{\bf R}_{j})-({\bf R}_j, \ddot{\bf R}_{i})=-2\frac{g_i-g_j}{g_i+g_j}(\dot{\bf R}_i, \dot{\bf R}_j)+2\sigma\epsilon_{ijk}a_k. \label{dt34} 
\end{eqnarray}
According to (\ref{dt25}), the equation (\ref{dt33}) is a consequence of $({\bf R}_i, {\bf R}_j)=\delta_{ij}$, and can be omitted from the system. 
Further, contracting Eq. (\ref{dt34}) with $-\frac12\epsilon_{kij}$ and using (\ref{dt31}) we get (see \cite{AAD23} for details)
\begin{eqnarray}\label{dt35}
I\dot{\boldsymbol\Omega}=[I{\boldsymbol\Omega},{\boldsymbol\Omega}]-\sigma[R^T{\bf k}, {\bf z}].  
\end{eqnarray}
The equation (\ref{dt31}) can  be rewritten in a form of first-order equation for determining $R_{ij}$
\begin{eqnarray}\label{dt36} 
\dot R_{ij}=-\epsilon_{jkm}\Omega_k R_{im}.    
\end{eqnarray}
For the latter use we enumerate the identities
\begin{eqnarray}\label{dt36.1}
R^T{\bf k}=k_i{\bf G}_i, \qquad (k_i {\bf G}_i)_j=({\bf k}, {\bf R}_j), \qquad (R{\bf z})_i=({\bf G}_i, {\bf z}), \qquad 
(R^T{\bf k}, {\bf z})=({\bf k}, R{\bf z})=k_i({\bf G}_i, {\bf z}). 
\end{eqnarray}

By construction, the system (\ref{dt35})-(\ref{dt36}) is equivalent to the second-order equations of motion (\ref{dt29}) for the rotation matrix $R_{ij}$. To exclude $\sigma$, we calculate second derivative of the constraint $({\bf k}, {\bf y}_c+R{\bf z})=0$: $({\bf k}, \ddot{\bf y}_c+\ddot R{\bf z})=0$ and use in this expression Eqs. (\ref{dt35}), (\ref{dt36}) as well as $\mu\ddot{\bf y}_c=-a\mu{\bf k}+\sigma{\bf k}$. After tedious calculations we obtain $\sigma$ through the physical variables as follows
\begin{eqnarray}\label{dt37}
\sigma(R, {\boldsymbol\Omega})=\left(\frac1\mu+({\bf a}, I{\bf a})\right)^{-1}\left\{a+([I{\boldsymbol\Omega}, {\boldsymbol\Omega}],I^{-1}[R^T{\bf k}, {\bf z}])+
([{\boldsymbol\Omega}, R^T{\bf k}], [{\boldsymbol\Omega}, {\bf z}])\right\} \qquad \qquad \qquad \cr 
\equiv\left(\frac1\mu+({\bf a}, I{\bf a})\right)^{-1}\left\{a+\frac{(I{\boldsymbol\Omega}, {\bf z})(I^2{\boldsymbol\Omega}, R^T{\bf k})-
(I{\boldsymbol\Omega}, R^T{\bf k})(I^2{\boldsymbol\Omega}, {\bf z})}{\det I}-
({\boldsymbol\Omega}, R^T{\bf k})({\boldsymbol\Omega}, {\bf z})+{\boldsymbol\Omega}^2(R^T{\bf k}, {\bf z})\right\}.
\end{eqnarray}
We recall our notation: ${\bf a}=I^{-1}[R^T{\bf k}, {\bf z}]$. 

Let us resume the obtained results. Heavy spinning top, with the base ${\bf y}_1(t)$ confined to move freely on a horizontal plane $\pi$ with unit normal vector ${\bf k}$, can be described in terms of orthogonal rotation matrix $R_{ij}(t)$, vector of angular velocity in the body $\Omega_i(t)$ and the position vector of center-of-mass ${\bf y}_c(t)$. They obey the following equations of motion
\begin{eqnarray}\label{dt38}
I\dot{\boldsymbol\Omega}=[I{\boldsymbol\Omega},{\boldsymbol\Omega}]-\sigma(R, {\boldsymbol\Omega})[R^T{\bf k}, {\bf z}],  
\end{eqnarray}
\begin{eqnarray}\label{dt39} 
\dot R_{ij}=-\epsilon_{jkm}\Omega_k R_{im}, \qquad \mbox{or, equivalently} \quad \dot{\bf G}_i=-[{\boldsymbol\Omega}, {\bf G}_i],  
\end{eqnarray}
\begin{eqnarray} 
\ddot{\bf y}_c=-\left(a-\frac{\sigma(R, {\boldsymbol\Omega})}{\mu}\right){\bf k}, \label{dt40} \\  ({\bf k}, {\bf y}_c+R{\bf z})=0,  \label{dt41}
\end{eqnarray}
where ${\bf z}\equiv{\bf x}_1(0)$ is initial position of the base in the center-of-mass system, the function $\sigma(R, {\boldsymbol\Omega})$ is given in Eq. (\ref{dt37}), $a$ is acceleration of gravity and $\mu$ is total mass of the spinning top. The initial conditions for the problem are $R_{ij}(0)=\delta_{ij}$, $\Omega_i(0)=\Omega_{0 i}$\,=const. This implies that any solution $R_{ij}(t)$ will be orthogonal matrix at any $t$. Given a solution to the problem, the evolution of the body's point ${\bf y}_N(t)$ with respect to Laboratory can be obtained using Eq. (\ref{dt19.1}): ${\bf y}_N(t)={\bf y}_c(t)+R(t){\bf x}_N(0)$, where  ${\bf x}_N(0)={\bf y}_N(0)-{\bf y}_c(0)$. 

The problem (\ref{dt38}), (\ref{dt39}) admits two integrals of motion, see below. 

Laboratory system was adapted with initial position of a spinning top such that the inertia tensor in Eq. (\ref{dt38}) has a diagonal form. 
Since the final equations  are composed of vectors and tensors, they preserve their form under rotations of the Laboratory system. However, after the rotation the diagonal matrix $I_i\delta_{ij}$ in Eq. (\ref{dt38}) generally turn into a symmetric matrix $I_{ij} $with non-zero off-diagonal elements. 

We conclude this section with some relevant comments. \par 

\noindent {\bf 1.}  The translational variable ${\bf y}_c$ does not enter into the equations (\ref{dt38}) and (\ref{dt39}) for rotational degrees of freedom $R_{ij}$ and $\Omega_i$, which therefore can be solved independently from the equations (\ref{dt40}) and (\ref{dt41}). Acording to Eq. (\ref{dt40}), the projection $N({\bf k}){\bf y}_c$ of the point ${\bf y}_c$ on the plane $\pi$ moves iniformly along a straight line. Given solution $R_{ij}$, $\Omega_i$ and $N({\bf k}){\bf y}_c$, the altitude $({\bf k}, {\bf y}_c)$ of this point  can be found algebraically from the constraint (\ref{dt41}). \par

\noindent {\bf 2.} The equations (\ref{dt38}) and (\ref{dt39}) of dancing spinning top look like the Euler-Poisson equations of a heavy body with fixed point: $I\dot{\boldsymbol\Omega}=[I{\boldsymbol\Omega},{\boldsymbol\Omega}]+a\mu[R^T{\bf k}, {\bf z}]$ 
and $\dot R_{ij}=-\epsilon_{jkm}\Omega_k R_{im}$, but with the acceleration of gravity being a complicated function $\sigma(R, {\boldsymbol\Omega})$ of rotational degrees of freedom. 

\par

\noindent {\bf 3.}  In Eq. (\ref{dt40}) for the center of mass ${\bf y}_c$, the reaction of supporting plane $\pi$  also looks like a fictitious gravitational field, shifting the gravitational acceleration from $a$ to $a-\sigma/\mu$.

\par

\noindent {\bf 4.}  For a body with one fixed point, the inertia tensor is calculated in the coordinate system with the origin at fixed point, while in the equations  (\ref{dt38}) and (\ref{dt39}) of dancing spinning top it must be calculated in the center-of-mass system.
\par

\noindent {\bf 4.}  From (\ref{dt40}) it follows that the reaction of supporting plane changes trajectory of the center-of-mass  in the gravity field. It can even balance the force of gravity: if $\sigma=a\mu$, and the initial  fall velocity of center of mass was zero, then it will be at the same height above the plane during the top's moviment.

\section{Symmetrical dancing top, an example of analytical solution.}

Consider a symmetrical top ($I_1=I_2\ne I_3$), whose base lies on the axis of inertia ${\bf R}_3$. The location of all vectors of this problem at initial instant $t=0$ is shown in the Figure \ref{DT_2}(b). The vectors ${\bf k}$, ${\bf R}_3(0)$ and ${\bf z}={\bf x}_1(0)$ were chosen in the plane of paper sheet, with ${\bf R}_3(0)$ in opposite direction with ${\bf x}_1(0)$. For the symmetrical top the choice of direction of inertia axes ${\bf R}_1(0)$ and ${\bf R}_2(0)$ in the plane orthogonal to ${\bf R}_3(0)$ is a matter of agreement \cite{AAD23_8,AAD23_5}. We take ${\bf R}_1(0)$ to be orthogonal to the plane of paper sheet. With this choice, the initial position of center-of-mass is ${\bf z}=(0, 0, -L)$, where $L$ is the distance between center of mass and fulcrum, while the unit normal vector to the supporting plane is ${\bf k}=(0, k_2, k_3)$. Then
\begin{eqnarray}\label{dt42}
R^T{\bf k}\equiv {\bf K}=\left(
\begin{array}{ccc}
k_2R_{21}+k_3R_{31} \\
k_2R_{22}+k_3R_{32} \\
k_2R_{23}+k_3R_{33} 
\end{array}\right), \quad 
[R^T{\bf k}, {\bf z}]=L\left(
\begin{array}{ccc}
-K_2 \\
K_1 \\
0 
\end{array}\right), \quad 
[{\boldsymbol\Omega}, {\bf z}]=L\left(
\begin{array}{ccc}
-\Omega_2 \\
\Omega_1 \\
0 
\end{array}\right), 
\end{eqnarray}
and the Euler equations acquire the following form 
\begin{eqnarray}\label{dt43} 
\dot\Omega_1=\frac{I_2-I_3}{I_2}\Omega_3\Omega_2+\frac{\sigma L}{I_2}K_2, \qquad \dot\Omega_2=-\frac{I_2-I_3}{I_2}\Omega_3\Omega_1-
\frac{\sigma L}{I_2}K_1, \qquad \dot\Omega_3=0. 
\end{eqnarray}
Hence for a symmetrical dancing top the third component of angular velocity represents one more integral of motion 
\begin{eqnarray}\label{dt44}
\Omega_3=\mbox{const}. 
\end{eqnarray}
Explicit form of the function $\sigma(R, {\boldsymbol\Omega})$ in these equations is
\begin{eqnarray}\label{dt45}
\sigma=\left[\frac1\mu+\frac{L^2}{I_2}(K_1^2+K_2^2)\right]^{-1}L\left\{\frac aL+\frac{I_3}{I_2}\Omega_3
(\Omega_1 K_1+\Omega_2 K_2)-(\Omega_1^2+\Omega_2^2)K_3. \right\}. 
\end{eqnarray}

Let us see whether there exists a solution to the problem (\ref{dt38})-(\ref{dt41}) for which the center of mass is always at the same height above the plane. For such a solution, the equation (\ref{dt40}) implies, that $\sigma$ should be a constant 
\begin{eqnarray}\label{dt46}
\sigma(R, {\boldsymbol\Omega})=a\mu, 
\end{eqnarray}
so we have a chance to find an exact analytical solution to a while problem. Besides, from Eq. (\ref{dt41}) we get  
$({\bf k}, R{\bf z})=(R^T{\bf k}, {\bf z})=-(({\bf k}, {\bf y}_c)=\mbox{const}$. According to Figure \ref{DT_2}(b), this means that point of base ${\bf x}_1(t)$ precesses around ${\bf k}$. As ${\bf R}_3(t)$ and ${\bf x}_1(t)$ lie on the same straight line, we expect that this motion be described by a rotation matrix consisting of the product of rotations around the axes ${\bf k}$ and ${\bf e}_3$:  $R(t, \alpha, \gamma)=R_{{\bf k}}(t, \gamma)\times R_{{\bf e}_3}(t, \alpha)$.  Therefore, we will look for a solution in the following form (see \cite{AAD23_5} for details):
\begin{eqnarray}\label{dt47}
R_{ij}=\left(
\begin{array}{ccc}\label{dt47.1}
\cos \gamma t\cos\alpha t-k_3\sin \gamma t\sin\alpha  t & -\cos \gamma t\sin\alpha  t-k_3\sin \gamma t\cos\alpha  t  &  k_2\sin \gamma t  \\
{} & {} & {} \\
k_3 \sin \gamma t\cos\alpha  t+(k_2^2 +k_3^2\cos \gamma t)\sin\alpha  t & 
-k_3 \sin \gamma t\sin\alpha  t+(k_2^2 +k_3^2\cos \gamma t)\cos\alpha  t & k_2k_3(1-\cos \gamma t) \\
{} & {} & {} \\
-k_2 \sin \gamma t\cos\alpha  t+k_2 k_3(1-\cos \gamma t)\sin\alpha  t &
k_2 \sin \gamma t\sin\alpha  t+k_2 k_3(1-\cos \gamma t)\cos\alpha  t & k_3^2 +k_2^2\cos \gamma t
\end{array}\right) 
\end{eqnarray}
where $\alpha$ is frequency of the top's rotation while $\gamma$ is frequency of the fulcrum precession. For the positive values of the frequences,  $R_{{\bf e}_3}(t, \alpha)$ is a clockwise rotation while $R_{{\bf k}}(t, \gamma)$ is a counter-clockwise.  Substituting this matrix into Eqs. (\ref{dt42}) and into (\ref{dt39}) written in the 
form: $\Omega_i=-\frac12\epsilon_{ijk}(R^T\dot R)_{jk}$,  we get
\begin{eqnarray}\label{dt48}
\Omega_1=\gamma k_2\sin\alpha t, \quad \Omega_2=\gamma k_2\cos\alpha t, \quad \Omega_3=\alpha+k_3\gamma, \label{dt48} \\
\quad K_1=k_2\sin\alpha t, \quad K_2=k_2\cos\alpha t, \quad K_3=k_3, \label{dt48.2}\qquad 
\end{eqnarray}
then $K_1^2+K_2^2=k_2^2$, $\Omega_1^2+\Omega_2^2=\gamma^2k_2^2$ and $\Omega_1 K_1+\Omega_2 K_2=\gamma k_2^2$. Using these equalities in (\ref{dt45}), we confirm that $\sigma$ is indeed a constant
\begin{eqnarray}\label{dt49}
\sigma=\frac{\mu L }{I_2+\mu L^2k_2^2}\left(\frac{I_2 a}{L}-(I_2-I_3)k_2^2k_3\gamma^2+I_3k_2^2\alpha\gamma\right). 
\end{eqnarray}
With $R_{ij}(t)$ and $\Omega_i(t)$ given by (\ref{dt47.1}) and (\ref{dt48}), the Poisson equations are satisfied by construction of these variables. Let us see 
whether $\Omega_i(t)$ obey the Euler equations. Substituting (\ref{dt48}) into (\ref{dt43}) we get
\begin{eqnarray}\label{dt50}
\alpha=\frac{I_2-I_3}{I_3}k_3\gamma+\frac{\sigma L}{I_3\gamma}.
\end{eqnarray}
Using (\ref{dt49}) in this expression, we exclude $\sigma$, obtaining final form of the function $\alpha(\gamma)$, that guarantees that the variables (\ref{dt48}) obey the Euler equations
\begin{eqnarray}\label{dt51}
\alpha=\frac{(I_2-I_3)k_3}{I_3}\gamma+\frac{\mu La}{I_3\gamma}. 
\end{eqnarray}
Substituting this function into (\ref{dt49}) we get $\sigma=a\mu$, that is the condition (\ref{dt46}) of self-consistency of our solution is satisfied. 

\begin{figure}[t] \centering
\includegraphics[width=10cm]{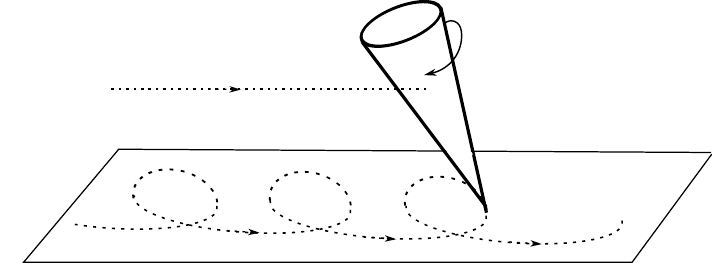}
\caption{The solution (\ref{dt47}), (\ref{dt51}) describes the movement of a symmetrical dancing top. Dotted lines show trajectories of center of 
mass and the fulcrum. An ice skater who decides to realize this figure with inclination of 45${}^\circ$, must develop the rotation speed near 25 rev./s.}\label{DT_3}
\end{figure}

Henceforce the rotation matrix (\ref{dt47.1}) with rotation frequency $\alpha$ given in Eq. (\ref{dt51}) and arbitrary precession frequency $\gamma$ describes the motion of a dancing spinning top, whose center of mass keep constant height above the supporting plane, while its fulcrum precesses on the plane around the vector ${\bf k}$, see Figure \ref{DT_3}. 

In the limiting case of vertically launched  top, $k_3=1$ and $k_2=0$, the solution (\ref{dt47.1}) turn into 
\begin{eqnarray}\label{dt52}
R_{ij}=\left(
\begin{array}{ccc}
\cos\varphi t & -\sin\varphi t & 0 \\
\sin\varphi t & \cos\varphi t & 0 \\
0 & 0 & 1 
\end{array}\right), 
\end{eqnarray}
with unique frequency $\varphi=\alpha+\gamma$.  This corresponds to a stationary rotation around fixed third inertia axis ${\bf R}_3(t)={\bf k}$.

The relationship (\ref{dt51}) coincide with that of Lagrange top \cite{AAD23_9}, but now the moments of inertia should be computed with respect to center of mass. The function $\alpha(\gamma)$ depends on the top's configuration $C=(I_2-I_3)/I_3$ and inclination $k_3$. To discuss this function, consider a conical top of height $h$, radius $r$ and the precession frequency $\gamma>0$. Its characteristics are \cite{Landau_8}
\begin{eqnarray}\label{dt53}
L=\frac{3h}{4}, \qquad I_1=I_2=\frac{3}{20}\mu(r^2+\frac{h^2}{4}), \qquad I_3=\frac{3}{10}\mu r^2.
\end{eqnarray}
We get $I_2-I_3=\frac{3\mu}{20}[\frac{h^2}{4}-r^2]$. So $C>0$ when $h>2r$ (high top), $C=0$ when $h=2r$ (totally symmetric top, $I_1=I_2=I_3$ ), and $C<0$ when $h<2r$ (low top).

Then the following cases arise. 

\begin{figure}[t] \centering
\includegraphics[width=09cm]{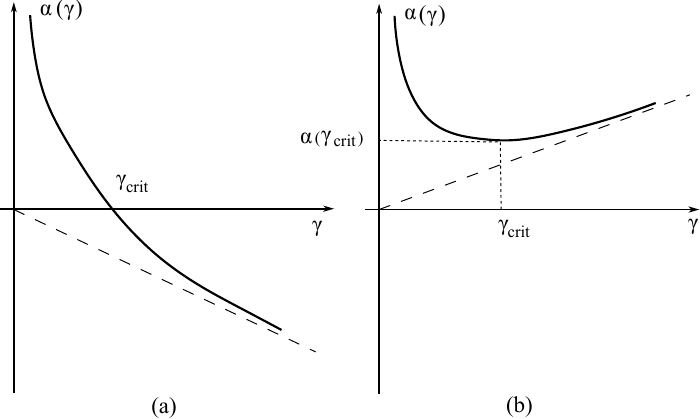}
\caption{The relationship between two frequences. (a) Low top located above the supporting plane, (b) High top located above the supporting plane.}\label{frequency}
\end{figure}

{\bf A.} The low top located above the supporting plane, that is 
\begin{eqnarray}\label{dt54} 
h<2r, \qquad \mbox{then} \quad C<0; \qquad k_3>0.
\end{eqnarray}    
The graph of the function $\alpha(\gamma)$ is drawn in Figure \ref{frequency}(a), and implies the following behavior of the top.

\noindent {\bf A1.} Slow precession of the fulcrum, $\gamma<\gamma_{crit}=\sqrt{\mu La/(I_2-I_3)k_3}$,  implies a clockwise rotation of the top around ${\bf R}_3(t)$. \par

\noindent {\bf A2.} When $\gamma=\gamma_{crit}$, the fulcrum precesses without rotation of the top around ${\bf R}_3(t)$. \par

\noindent {\bf A3.} Rapid precession of the fulcrum, $\gamma>\gamma_{crit}$, implies counter-clockwise rotation of the top around ${\bf R}_3(t)$. In the limit $\gamma\rightarrow\infty$ the rotation frequency $|\alpha|$  grows linearly with $\gamma$. 

{\bf B.} The high top located below the supporting plane 
\begin{eqnarray}\label{dt55} 
h>2r, \qquad \mbox{then} \quad C>0; \qquad k_3<0,
\end{eqnarray}
has a similar bevavior. 

{\bf C.} The high top located above the supporting plane 
\begin{eqnarray}\label{dt56} 
h>2r, \qquad \mbox{then} \quad C>0; \qquad k_3>0.
\end{eqnarray}
The graph of the function $\alpha(\gamma)$ is drawn in Figure \ref{frequency}(b), and implies the following behavior of the top.

\noindent {\bf C1.} The top rotating around ${\bf R}_3(t)$ with the frequency less than 
\begin{eqnarray}\label{dt56.1} 
\alpha(\gamma_{crit})=2\sqrt{(I_2-I_3)\mu Lak_3}/I_3,  
\end{eqnarray}
cannot maintain a constant height of the center of mass above the supporting plane.

\noindent {\bf C2.} There is only one rotation frequency $\alpha(\gamma_{crit})$ at which the fulcrum precesses with the 
frequency $\gamma_{crit}=\sqrt{\mu La/(I_2-I_3)k_3}$.

\noindent {\bf C3.} For each rotation frequency $\alpha>\alpha(\gamma_{crit})$, there are two possible precession frequences of the fulcrum at which the center of mass maintain a constant height above the supporting plane
\begin{eqnarray}\label{dt57} 
\gamma= \frac{\alpha I_3\pm\sqrt{\alpha^2I_3^2-4(I_2-I_3)\mu Lak_3}}{2(I_2-I_3)k_3}. 
\end{eqnarray}  

Let's assume that an ice skater decided to realize the figure "dancing top" with an inclination equal to $k_3=\cos\frac{\pi}{4}$.
The minimum rotation speed, required to perform this figure, must be $\alpha(\gamma_{crit})$ given in Eq. (\ref{dt56.1}). Approximating the skater with a conical top with $\mu=70$ kg, $h=1.7$ m, $r=0,2$ m and using (\ref{dt56.1}), we 
get $\alpha(\gamma_{crit})\approx 158, 45 ~\mbox{rad/s}\approx 25 ~ \mbox{rev./s}$. If he could develop such a speed of rotation, his precession frequency on ice field would 
be $\gamma_{crit}=\sqrt{\mu La/(I_2-I_3)k_3}\approx 2,1 ~ \mbox{rev./s}$.

{\bf D.} The low top located below the supporting plane 
\begin{eqnarray}\label{dt58} 
h<2r, \qquad \mbox{then} \quad C<0; \qquad k_3<0,
\end{eqnarray}
has a similar bevavior. 

{\bf E.} For the totally symmetric conical top  
\begin{eqnarray}\label{dt59} 
h=2r, \qquad \mbox{then} \qquad I_1=I_2=I_3, \qquad  C=0, \qquad \alpha=\frac{b}{I_3\gamma},
\end{eqnarray}
the relationship between frequncies of rotation and precession does not depend on inclination. The fulcrum of slow spinning top  must precess at high speed on the supporting plane. A fast spinning top precesses slowly.

The same behavior has launched horizontally ($k_3=0$) symmetric top. In this case the relationship between frequncies of rotation and precession does not depend on the top's geometry $C$.

\section{Integrability of equations of a dancing spinning top.} 

{\bf Integration with use of Hamiltonian vector field.} For the first-order system (\ref{dt38}) and (\ref{dt39}), we can use the known formula of Hamiltonian mechanics \cite{AAD_2022, AAD23_2} to write its general solution through exponential of Hamiltonian vector field. Denoting the r.h.s. of equations (\ref{dt38}) and (\ref{dt39}) as follows
\begin{eqnarray}\label{dt60}
[I^{-1}I{\boldsymbol\Omega},{\boldsymbol\Omega}]_i-
\sigma(R, {\boldsymbol\Omega})(I^{-1}[R^T{\bf k}, {\bf z}])_i\equiv H_i(R, {\boldsymbol\Omega}), \qquad   
-\epsilon_{jkm}\Omega_k R_{im}=H_{ij}(R, {\boldsymbol\Omega}), 
\end{eqnarray}
the solution reads
\begin{eqnarray}\label{dt61}
R_{ij}(t, R_{0 kp}, \Omega_{0 k})=e^{tH_{kp}(R_0, {\boldsymbol\Omega}_0)\frac{\partial}{\partial R_{0 kp}}+tH_k(R_0, {\boldsymbol\Omega}_0)\frac{\partial}{\partial\Omega_{0 k}}}R_{0ij}. 
\end{eqnarray}
After applying the differential operator in the exponential, $R_{0 kp}=\delta_{kp}$ should be substituted in each term of the obtained power series.
The resulting function $R_{ij}(t, \Omega_{0 k})$ will represent the general solution to the problem with three integration constants $\Omega_{0k}$.

{\bf Integrals of motion.}  According to (\ref{dt7}),  the projection of center-of-mass velocity on the plane $\pi$ gives two independent integrals of motion: 
\begin{eqnarray}\label{dt16.0}
N({\bf k})\dot{\bf y}_c\equiv{\bf V}_c=\mbox{const}, \qquad \mbox{let us denote} \quad E_c\equiv\frac12\mu[N({\bf k})\dot{\bf y}_c]^2.  
\end{eqnarray}
They are related with the translational part (\ref{dt40}) and (\ref{dt41}) of our equations of motion.

Taking scalar product of the equations for $\ddot{\bf x}_N$, $\ddot{\bf y}_c$ of the system (\ref{dt13}) with $\dot{\bf x}_N$, $\dot{\bf y}_c$,  and then summing all them,  all terms with $\lambda$ vanish, and we obtain that total energy 
\begin{eqnarray}\label{dt16}
E'=\frac12\mu\dot{\bf y}_c^2+\frac12 \sum_{N=1}^nm_{N}\dot{\bf x}_N^2+a\mu({\bf k}, {\bf y}_c),  
\end{eqnarray}
is preserved along any true trajectory of the body: $\frac{dE'}{dt}=0$. It contains the integral of motion $E_c$, let us separate it. Using the 
decomposition $\dot{\bf y}_c^2=[N\dot{\bf y}_c+P\dot{\bf y}_c]^2=[N\dot{\bf y}_c]^2+({\bf k}, \dot{\bf y}_c)^2$ we rewrite (\ref{dt16}) as follows
\begin{eqnarray}\label{dt16.1}
\frac12 \sum_{N=1}^nm_{N}\dot{\bf x}_N^2+\frac12\mu({\bf k}, \dot{\bf y}_c)^2+a\mu({\bf k}, {\bf y}_c)=E'-E_c\equiv E. 
\end{eqnarray}
Using the constraint $({\bf k}, {\bf y}_c+R{\bf z})=0$ we can excude the center-of-mass variable from this expression
\begin{eqnarray}\label{dt16.1}
\frac12 \sum_{N=1}^nm_{N}\dot{\bf x}_N^2+\frac12\mu({\bf k}, \dot R{\bf z})^2-a\mu({\bf k}, R{\bf z})=E. 
\end{eqnarray}
Further, using Eqs. (\ref{dt36}) and (\ref{dt36.1}) we get the final form of this integral of motion
\begin{eqnarray}\label{dt16.2}
\frac12 I_i(\Omega_i)^2+\frac12\mu({\boldsymbol\Omega}, [R^T{\bf k}, {\bf z}])^2 -a\mu(R^T{\bf k}, {\bf z})=E. 
\end{eqnarray}
 
Taking  vector product of the equations for $\ddot{\bf x}_N$ of the system (\ref{dt13}) with ${\bf x}_N$, and then summing all them, we get equation for angular momentum of the body with respect to the center of mass 
\begin{eqnarray}\label{dt17}
\frac{d}{dt}{\bf m}=\sigma [{\bf k}, {\bf x}_1], \qquad \mbox{where} \qquad {\bf m}=\sum_{N=1}^{n}m_N[{\bf x}_N, \dot{\bf x}_N]. 
\end{eqnarray}
Its projection on the direction of ${\bf k}$ gives one more integral of motion
\begin{eqnarray}\label{dt18}
\frac{d}{dt}({\bf k}, {\bf m})=0, \qquad \mbox{or} \quad \frac{d}{dt}({\bf k}, RI{\boldsymbol\Omega})=0. 
\end{eqnarray}
For a symmetrical top, we have found one more integral of motion (\ref{dt44}).

{\bf Discussion of integrability according to Liouville.} The Euler-Poisson equations (\ref{dt38}) and (\ref{dt39}) represent the Hamiltonian system \cite{AAD23} written for six independent phase-space degrees of freedom (three independent coordinates contained in the orthogonal matrix $R^TR={\bf 1}$, and three momenta $\Omega_i$). For the case of a symmetrical dancing top, the system admits three integrals of motion $T_i=(E, ({\bf k}, {\bf m}), \Omega_3)$. According to Liouville's theorem \cite{Arn_1,Fom_2004}, if they are in involution with respect to canonical Poisson brackets: $\{T_i, T_j\}=0$, the problem can be reduced to calculation of integrals of known functions and some algebraic operations. To confirm the Liouville's integrability, we need to construct Hamiltonian formulation of the theory (\ref{dt22}) in terms of a set of independent variables subject to canonical Poisson brackets. For the case of a free rigid body this this can be achieved \cite{AAD23_1}, using the parameterization of rotation matrix in terms of unconstrained vector $n_i$ as follows
\begin{eqnarray}\label{dt70}
R_{ij}({\bf n})=\frac{1}{1+{\bf n}^2}\left[(1-{\bf n}^2)\delta_{ij}+2n_i n_j-2\epsilon_{ijk}n_k\right]. 
\end{eqnarray}
Compared to the free theory, the Lagrangian (\ref{dt22}) contains an interaction term and one more constraint. In terms of $n_i, y_c^j$ and their conjugate momenta $\pi_i, p_c^j$, the corresponding Hamiltonian reads
\begin{eqnarray}\label{dt71}
H=\frac18 I^{-1}_{ij}(\tilde A{\boldsymbol\pi})_i(\tilde A{\boldsymbol\pi})_j+\frac{1}{2\mu}{\bf p}_c^2-\sigma({\bf k}, {\bf y}_c+
R({\bf n}){\bf z})+a\mu({\bf k}, {\bf y}_c),
\end{eqnarray}
where $\tilde A_{ij}=\delta_{ij}+n_j n_j+\epsilon_{ijk}n_k$. This leads to the Hamiltonian theory with two second-class 
constraints, $({\bf k}, {\bf y}_c+R({\bf n}){\bf z})=0$ and $\frac{1}{\mu}({\bf k}, {\bf p}_c)+({\bf k}, {\bf B}({\bf n}, {\boldsymbol\pi}){\bf z})=0$, 
where the Poisson bracket ${\bf B}=\{R{\bf z}, H\}$ is a function of canonical variables $n_i$ and $\pi_j$. To construct the physical Hamiltonian without the auxiliary variable $\sigma$, we need to construct the Dirac bracket that take into account these two constraints. In the resulting theory the brackets  $\{n_i, n_j\}_{DB}$, $\{\pi_i, \pi_j\}_{DB}$ and  
$\{\pi_i, n_j\}_{DB}$ no longer have a canonical form, and therefore the Liouville's integrability is under the question.

\begin{acknowledgments}
I am grateful to Sergei M. Kuzenko for stimulating discussions of the Kovalevskaya top.
The work has been supported by the Brazilian foundation CNPq (Conselho Nacional de
Desenvolvimento Cient\'ifico e Tecnol\'ogico - Brasil).
\end{acknowledgments}

\end{document}